\documentclass{appolb}
\usepackage{epsfig}

\usepackage{amsmath, amssymb}
\usepackage{slashed}
\usepackage{subfigure}
\usepackage{color}

\newcommand{\Phibar}{\ensuremath{\bar{\Phi}}}

\def\eq#1{(\ref{#1})}
\def\Eq#1{Eq.~(\ref{#1})}

\begin{document}  
% \eqsec  % uncomment this line to get equations numbered by (sec.num)
\title{The Impact of Fluctuations on QCD Matter%
  \thanks{Presented at the HIC for FAIR Workshop and XXVIII Max Born
    Symposium ``Three Days on Quarkyonic Island'', Wroclaw, May 19-21,
    2011.}%
} \author{Tina Katharina Herbst$^a$, Jan M. Pawlowski$^{b,c}$\and
  Bernd-Jochen Schaefer$^{a,d}$ \address{$^a$ Institut f\"ur Physik,
    Karl-Franzens-Universit\"at Graz, A-8010 Graz} \address{$^b$
    Institut f\"ur Theoretische Physik, Universit\"at Heidelberg,
    D-62910 Heidelberg} \address{$^c$ ExtreMe Matter Institute EMMI,
    GSI Helmholtzzentrum f\"ur Schwerionenforschung mbH, D-64291
    Darmstadt} \address{$^d$ Institut f\"ur Theoretische Physik,
    Justus-Liebig-Universit\"at Gie{\ss}en, D-35392 Gie{\ss}en} }

\maketitle
\begin{abstract}
  We study the effect of quantum and thermal fluctuations as well as
  the mass dependence of the phase structure of QCD at
  finite temperature and density within a dynamical
  Polyakov-loop--extended quark-meson model.  The glue dynamics is
  simulated by the Polyakov-loop potential, also including the
  back-coupling of the matter sector to the glue dynamics.  In the
  chiral limit, the chiral phase transition at large chemical
  potential and low temperature splits into two transition branches.
  For non-vanishing pion masses the chiral transition at small
  chemical potential changes from a phase transition to a
  crossover. We close with a discussion of a systematical improvement
  of the current model towards full QCD.

\end{abstract}
 %Finite-temperature field theory
% 12.38.Gc}		%Lattice QCD calculations (see also 11.15.Ha
                        %Lattice gauge theory)

%-----------------------------------------------------
%-----------------------------------------------------
\section{Introduction}
\label{sec:intro}
Over the past two decades, much progress has been made experimentally
and theoretically in our understanding of the phase structure of
strongly-interacting matter \cite{Friman:2011zz}. This has been
achieved in a combination of first principle continuum computations in
QCD, see e.g.\ \cite{Pawlowski:2010ht}, lattice simulations, see e.g.\
\cite{Philipsen:2011zx,Borsanyi:2010cj,Bazavov:2010sb}, as well as
computations in low energy effective models for QCD, see e.g.\
\cite{Fukushima:2010bq,Schaefer:2007pw,Megias:2004hj,
  Ratti:2005jh,Fukushima:2003fw,Dumitru:2003hp}. At vanishing chemical
potential it is established by now that there are two crossover
transitions at similar temperatures, see
e.g. \cite{Borsanyi:2010cj,Bazavov:2010sb,Braun:2009gm}.  Presently,
it is still much under debate whether the two transitions coincide
over the whole phase diagram or deviate at some chemical potential,
leave aside the phase structure at larger chemical potential.  In
lattice simulations progress is hampered by the sign problem at finite
chemical potential. The sign problem does not affect the continuum
approaches which, however, have to deal with the increase of relevant
degrees of freedom and fluctuations as well as the strong correlations
at increasing density.

In this context it has been argued that the QCD effective models can
be understood as well-controlled approximations of full first
principle functional approaches to QCD, see
\cite{Pawlowski:2010ht,Herbst:2010rf}. This connection allows to
successively constrain and finally determine the sensitive model
parameters, and hence to gain qualitatively on predictive power within
these models. In turn, the models provide a more direct access and
understanding of the mechanisms at work for e.g.\ the chiral and
confinement/deconfinement dynamics, see
e.g. \cite{Meisinger:1995ih,Braun:2011fw}. The Polyakov-extended
quark-meson (PQM) model \cite{Schaefer:2007pw} is specifically
well-adapted to the embedding into first principle approaches to
QCD. By now it is well-studied also beyond the mean-field level at
vanishing \cite{Skokov:2010wb} and finite density
\cite{Herbst:2010rf,Skokov:2010uh,Morita:2011eu,Skokov:2011ib}. In
this model the matter fluctuations are included directly within a
fully non-perturbative functional renormalization group (FRG)
approach, for reviews on QCD and QCD-effective models see
\cite{Litim:1998nf,Berges:2000ew,Pawlowski:2005xe,%
Gies:2006wv,Schaefer:2006sr,Braun:2011pp}.

In Polyakov-extended models (part of) the glue dynamics is included
with a Polyakov-loop potential in the free energy. This potential is
usually constrained with the expectation value and thermodynamics of
lattice Yang-Mills (YM) theory
\cite{Ratti:2005jh,Fukushima:2003fw}. Clearly, this does not
completely fix the potential and the phase structure is sensitive to
the chosen potential, see e.g.\ \cite{Schaefer:2009ui}. This
necessitates a better determination of the potential. Corrections due
to matter fluctuations have been computed in \cite{Schaefer:2007pw},
reviewed in \cite{Schaefer:2011pn}, and have been confirmed at
vanishing temperature with a direct comparison of the Polyakov-loop
potentials in Yang-Mills theory \cite{Braun:2007bx} and QCD
\cite{Braun:2009gm}.  Further work aiming at the determination also at
finite chemical potential is in progress. In summary, the PQM model
provides us with a good approximation of low energy QCD at not too
large chemical potential, and its direct relation to continuum QCD
approaches allows us to strengthen this relation successively.

In the present contribution we study the interrelation between chiral
symmetry breaking, its restoration at large temperatures and/or
chemical potential and the confinement/deconfinement transition within
the dynamical two-flavour PQM model. We map out the full phase diagram and also
provide a conservative estimate for the possible emergence of a
critical endpoint. Further insights into the mechanisms at work are
extracted by varying the explicit chiral symmetry breaking parameter,
and hence the Goldstone-boson (pion) mass.

%-----------------------------------------------------
%-----------------------------------------------------
\section{QCD flows and the matter back-coupling}
\label{sec:phase_structure}

The dynamical Polyakov-extended quark-meson model used in the present
contribution consists of a Yukawa-type model of propagating quarks and
mesons (pions $\vec \pi$ and sigma $\sigma$). As a low-energy
effective model of QCD it is formed dynamically when fully taking into
account the four-fermi interaction that arises from quark and gluon
fluctuations. The mesonic resonances that dynamically form at lower
energies, in particular in the $s$-channel of the four-fermi
interaction, lead to effective low-energy degrees of freedom, the
mesons. In the flow equation approach this is described by dynamical
hadronization, see
\cite{Gies:2001nw,Gies:2002hq,Pawlowski:2005xe,Floerchinger:2009uf}.
This approach to QCD has been put forward for QCD in
\cite{Braun:2009gm,Braun:2008pi} and is reviewed in \cite{Pawlowski:2010ht}.

\begin{figure}[t]
 \centering
 \includegraphics[width=.75\textwidth]{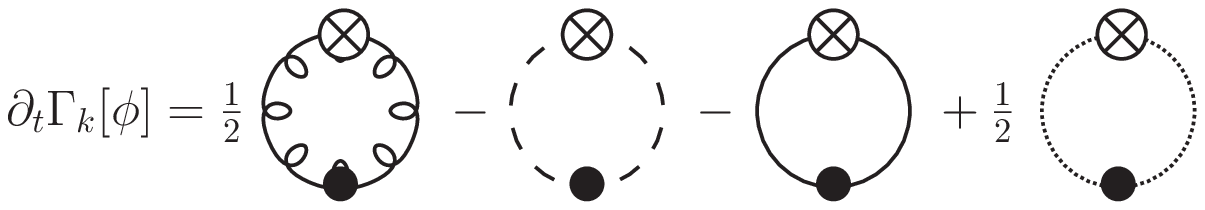}
 \caption{RG flow for QCD, including gluon, ghost, meson and quark
   contributions, respectively.}
 \label{fig:QCD_flow}
\end{figure}

The present dynamical PQM model is discussed in detail in
\cite{Herbst:2010rf}, including its direct connection to the FRG
approach to QCD. The flow equation for the free energy of QCD is shown
in Fig.~\ref{fig:QCD_flow}. It is important to emphasize that the
depicted flow is exact, no higher loop diagrams are missing.  Here,
$\Gamma_k[\phi]$ stands for the effective action in the presence of an
infrared cut-off scale $k$ and the free energy is given by the
effective action evaluated at the equation of motion. For each degree
of freedom a fully dressed one-loop structure emerges. For example,
the first two loops for the gluons and ghosts comprise the pure glue
sector in QCD: the full gluon and ghost propagators are screened in
the presence of dynamical quarks and mesons, respectively.
%%%%%%%%%%%%%%%%%%%%%%%%%%%%%%%%%%%%%%

\begin{figure}[t]
 \centering
  \includegraphics[width=.3\textwidth]{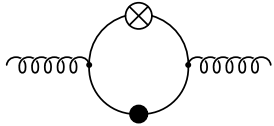} 
  \caption{Quark loop correction to the gluon propagator contributing
    to the matter back-coupling to the gauge sector present in full
    QCD.}
 \label{fig:backcoupling}
\end{figure}
%%%%%%%%%%%%%%%%%%%%%%%%%%%%%%%%%%%%%%
An example for
the matter contribution to the glue propagator is shown in
Fig.~\ref{fig:backcoupling} with the vacuum polarisation. The quark
loop screens the gluon propagator, and in particular with increasing
density the in-medium effects get larger. This has crucial
effects, such as the reduction of the deconfinement transition
temperature of the pure YM system when full QCD, including the
dynamical quarks and meson, is considered.

The simple additive structure has important consequences for the PQM
model. We can easily identify the different parts of the model in
terms of full QCD fluctuations. For example, the fluctuating
quark-meson sector of the model is given by the last two diagrams in
Fig.~\ref{fig:QCD_flow}. In other words, removing the glue sector in
Fig.~\ref{fig:QCD_flow} leaves us with a dynamical quark-meson model,
for a review see \cite{Schaefer:2006sr}. In turn, the glue and ghost
loops in Fig.~\ref{fig:QCD_flow}, evaluated in the background of a
Polyakov-loop $\Phi$, give the Polyakov-loop potential in full
QCD. This potential only agrees with the Yang-Mills potential if we
neglect the matter fluctuations, such as the vacuum polarisation
depicted in Fig.~\ref{fig:backcoupling}. This entails that with
Fig.~\ref{fig:backcoupling} we have access to the change of the
Polyakov-loop potential in the presence of matter fluctuations.

It is precisely this simple structure which has been used in
\cite{Schaefer:2007pw} for a phenomenological estimate of the matter
back-coupling to the glue sector.  On the practical level, the
back-coupling changes the dynamical Yang-Mills scale $\Lambda_{\rm
  YM}$, present in the gluon and ghost propagators, to the dynamical
QCD scale $\Lambda_{\rm QCD}$. In the Polyakov-loop potential this
scale manifests itself in the critical temperature $T_0$. Therefore,
in full QCD this parameter receives a flavour and chemical potential
dependence via the vacuum polarisation, see
Fig.~\ref{fig:backcoupling}, leading to $T_0(N_f,\mu)$
\cite{Schaefer:2007pw}. By now, we also have access to the full QCD
potential at vanishing chemical potential derived from solving the
fully coupled QCD flows
\cite{Pawlowski:2010ht,Braun:2009gm,Haas:2010bw}. This result confirms
the phenomenological estimate in \cite{Schaefer:2007pw}, and will soon
be extended to finite chemical potential, thus resolving the current
ambiguity in the Polyakov-loop potential in the Polyakov-loop extended
models.
 
The discussion above allows us to put forward the flow equation of QCD
where the gluonic degrees of freedom have been integrated out. This
leads to a flow for the free energy which only involves the last two
loops in Fig.~\ref{fig:QCD_flow} and the first two loops lead to the
Polyakov-loop glue potential $\Omega_{\rm glue}$. They also lead to
modifications of the matter interaction which is included in the initial
conditions in the quark-meson sector that capture the correct vacuum
physics. Ignoring the additional modification of the quark and meson
dispersion this leads us finally to the flow equation
\cite{Skokov:2010wb,Herbst:2010rf} for the QCD free energy $\Omega_{\rm
  QCD,k}=\Omega_k$, 
\begin{eqnarray}\label{eq:PQMflow}
  k\partial_k \Omega_k(\sigma,\vec \pi,\Phi,\bar\Phi) & = & \frac{k^5}{12\pi^2}
  \left\lbrace \frac{1}{E_\sigma}\coth\left(\frac{E_\sigma}{2T}\right) + \frac{3}{E_\pi}
    \coth\left(\frac{E_\pi}{2T}\right)\right.\\
  & &\left. -\frac{4N_cN_f}{E_q}\left[1 - N_q( T,\mu; \Phi,\Phibar) - N_{\bar q}
      ( T,\mu; \Phi,\Phibar)\right] \right\rbrace\,, \nonumber
\end{eqnarray} 
where $k$ is the infrared cut-off scale. The quasi-particle energies
are given by, $ i=q, \pi, \sigma$, 
\begin{eqnarray}\label{eq:quasienergies}
E_i = \sqrt{k^2+m_i^2}\,,\qquad m_q^2=h^2\phi^2\,,\quad  
m_\pi^2=2\Omega_k'\,, \quad m_\sigma^2=2\Omega_k'+4\phi^2\Omega''_k\,,
\end{eqnarray}
with $\phi=(\sigma,\vec\pi)$ and primes denote derivatives with
respect to $\phi^2$. The Polyakov-loop enhanced quark/anti-quark
occupation numbers read
\begin{eqnarray}
    N_{q}(T,\mu;\Phi,\bar{\Phi}) \!&\! = \!&\! \dfrac{1+2\bar{\Phi} e^{(E_q-\mu)/T}+\Phi
      e^{2(E_q-\mu)/T}}{1+3\bar{\Phi}
      e^{(E_q-\mu)/T}+3\Phi e^{2(E_q-\mu)/T}+e^{3(E_q-\mu)/T}}
\end{eqnarray}
and $N_{\bar q}(T,\mu;\Phi,\bar{\Phi}) \equiv
N_{q}(T,-\mu;\bar{\Phi},\Phi)$. \Eq{eq:PQMflow} as it stands is still
a flow equation for the free energy of full QCD with the approximation
of a classical dispersion of quarks and mesons. The gluons have been
integrated out, their dynamics is directly stored in the glue
Polyakov-loop potential $\Omega_{\rm glue}$ as well as the initial
conditions for $\Omega_\Lambda$. Hence, only the matter part $\Omega_{{\rm matter},k}$ depends on
the cut-off scale $k$, to wit $\partial_k\Omega_{\rm
  QCD,k}=\partial_k\Omega_{{\rm matter},k}$. This leads us to the final
expression for the QCD free energy $\Omega_{\rm QCD}=\Omega_{{\rm
    QCD},k=0}$, 
\begin{eqnarray}\nonumber 
  \Omega_{\rm QCD}(\sigma,\vec \pi,\Phi,\bar\Phi) &=& \Omega_{\rm glue}(\Phi,\bar\Phi)+
\int_\Lambda^0 dk\,\partial_k \Omega_{ {\rm matter},k}(\sigma,\vec \pi,\Phi,\bar\Phi)
\\[1ex]
&& +\Omega_{ {\rm matter},\Lambda}(\sigma,\vec \pi,\Phi,\bar\Phi)\,.
\label{eq:freeenergy}
\end{eqnarray} 
The flow equation \eq{eq:PQMflow} is indeed that of the matter sector
of the two-flavour QCD flow derived in \cite{Braun:2009gm} with
$\Phi=\Phi(A_0)$. At sufficiently large initial cut-off scale
$\Lambda$ the initial matter part of the free energy, $\Omega_{ \rm
  matter,\Lambda}$, is just a local Yukawa-type action of quarks and
mesons, its parameters are fixed with phenomenology in the vacuum with
$f_\pi$, $m_\pi$ and $m_\sigma$. We also remark that the independence
of the full free energy from the initial cut-off scale $\Lambda$, that
is $\partial_\Lambda \Omega_{\rm QCD}\equiv 0$, enforces
$\Lambda$ dependent terms in $\Omega_{ {\rm matter},\Lambda}$. These
terms can be determined from the flow at $\Lambda$, see the reviews
\cite{Litim:1998nf,Berges:2000ew,Pawlowski:2005xe,%
  Gies:2006wv,Schaefer:2006sr,Braun:2011pp}.  Phenomenologically, they
can be understood as the high-energy part of the vacuum fluctuations,
see \cite{Skokov:2010sf, Schaefer:2011ex}.

In the following we present results based on the full integration of
the flow of $\Omega_{\rm QCD}$ without any further approximation: the
flow \eq{eq:PQMflow} is solved on a $\sigma$-grid for general fixed
backgrounds $\Phi$, $\bar\Phi$. Note in this context that the
couplings in the initial condition $ \Omega_{\rm
    QCD,\Lambda}$ are insensitive to the choice of $\Phi$,
$\bar\Phi$. By mapping out the $\Phi$, $\bar\Phi$-plane this provides
us with the final result $\Omega_{\rm QCD}(\sigma,\vec
\pi,\Phi,\bar\Phi)=\Omega_{\rm QCD,k=0}(\sigma,\vec
\pi,\Phi,\bar\Phi)$. This goes beyond the approximation used in
\cite{Herbst:2010rf} where the flow was solved on the mean-field
solutions $\Phi(\sigma),\bar\Phi(\sigma)$. It has been argued there
that the latter already provides a quantitatively reliable
approximation to the full solution and the present results fully
confirm this argument. More details will be presented elsewhere
\cite{HPS}.  The present computation allows us to solve the equations
of motion for $\Phi$, $\bar\Phi$ and $\sigma$ without further
approximations and hence to determine the free energy and other
thermodynamic quantities.

The above discussion of the Polyakov loop potential only exemplifies how
results obtained within the first principle FRG approach to QCD can be
used in order to constrain model computations. It is by no means
restricted to the current example and provides a way of systematically
improving the models towards full QCD. Moreover, lattice results
provide further input and in particular also help to improve the
quantitative precision of the QCD flows. In our opinion, in
combination this opens a systematic path of mapping out the phase
diagram of QCD. 

\begin{figure}
  \centering
    \includegraphics[width=.49\textwidth]{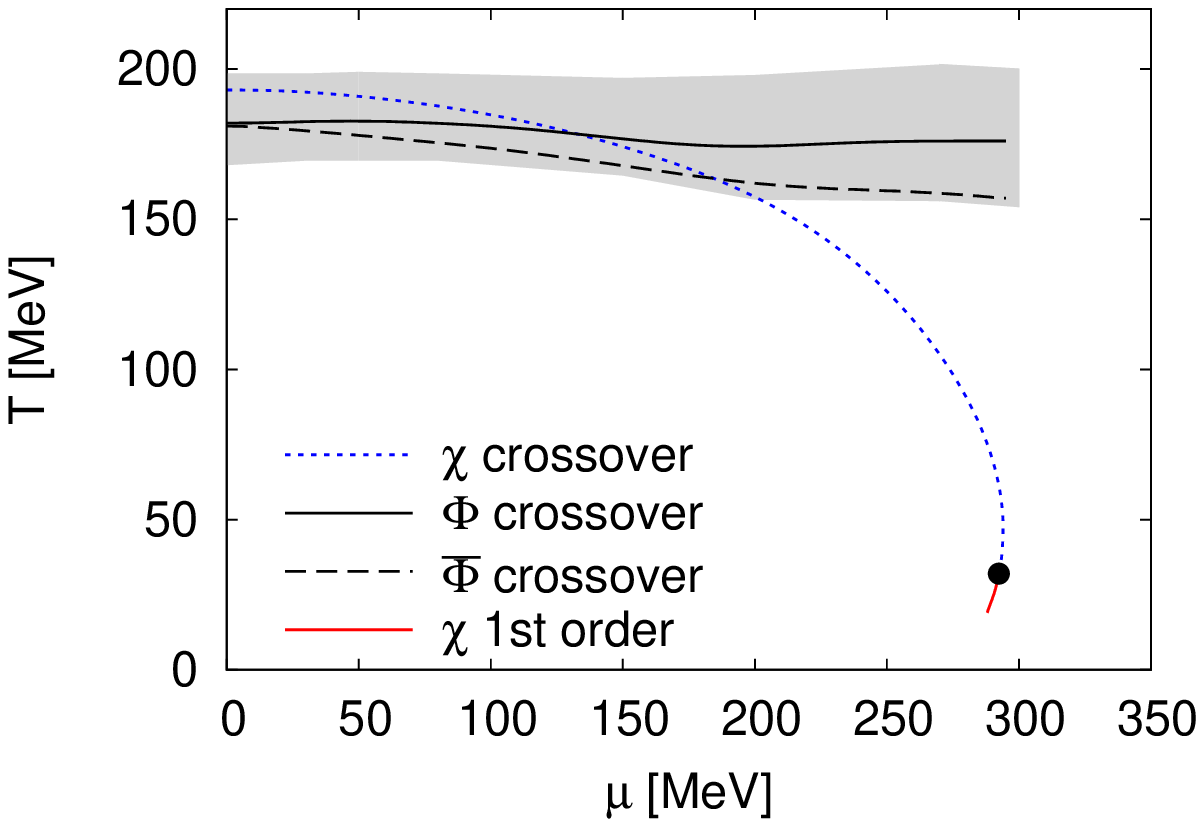}
    \includegraphics[width=.49\textwidth]{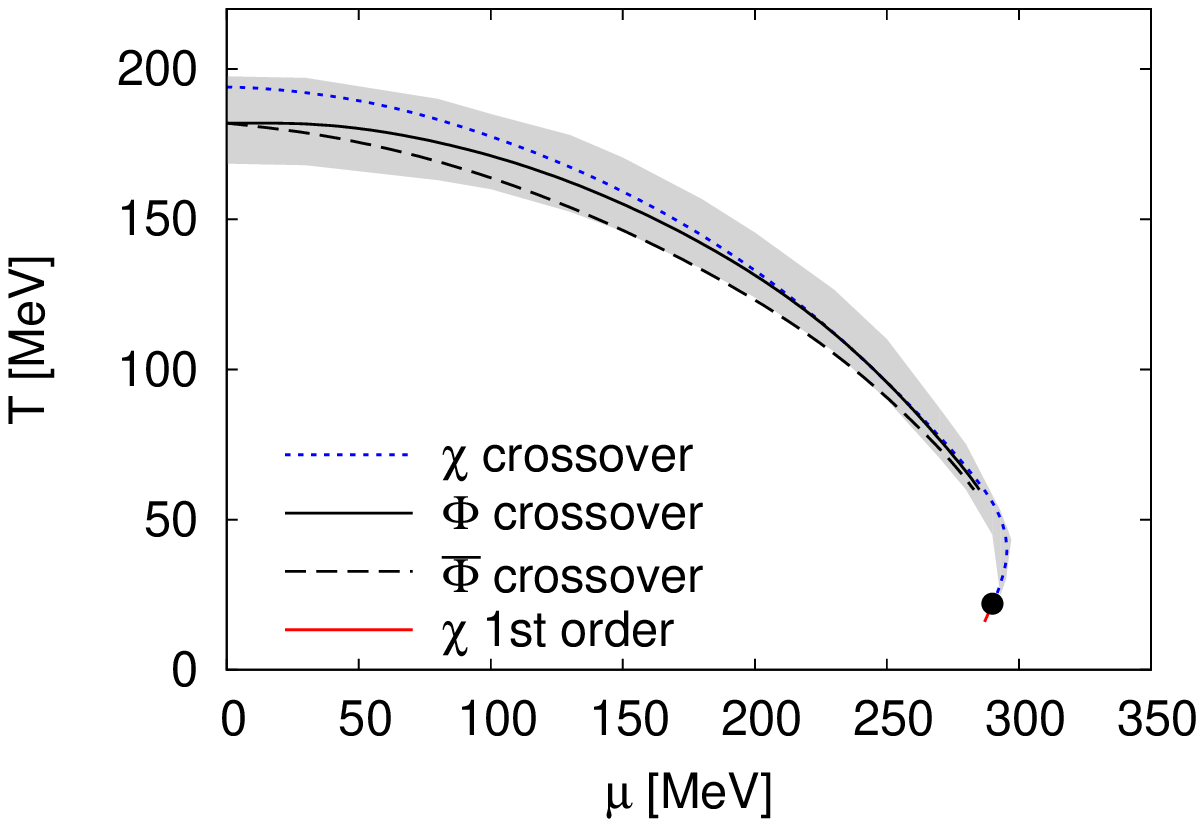}
    \caption{Phase diagram with constant $T_0$ (left) and with
    running $T_0(\mu)$ (right).} 
  \label{fig:phasediagram_phys}
\end{figure}

\section{Phase structure at physical pion masses}

Fig.~\ref{fig:phasediagram_phys} summarizes our findings for the
two-flavour phase structure at physical pion masses, $m_\pi=138$ MeV,
and for a constant $T_0$ (left) and running $T_0(\mu)$ parameter
(right). The gray band denotes the width of the temperature derivative
of the Polyakov-loop (solid black line) at $80\%$ of its maximum
height.  The dashed black line labels the inflection point of the
conjugate Polyakov-loop. Both Polyakov-loop transitions, in the
following referred to as deconfinement transition, are crossovers at
low $\mu$. With a constant $T_0$ (left panel), we find an almost
$\mu$ independent deconfinement temperature which is slightly below
the chiral crossover (dotted blue line).  At $\mu\approx 200$ MeV the
chiral and deconfinement transitions start to deviate and we find a
region in the phase diagram where chiral symmetry is partially
restored and confinement still persists. The picture changes when we
include the matter back-reaction to the gluonic sector via a running
$T_0(\mu)$ (right panel). In this case, both transitions lie close to
each other throughout the whole phase diagram and no room for a
chirally restored and confined phase is left. Remarkably, a similar
scenario was found in another recent two-flavour non-perturbative
functional study with Dyson-Schwinger equations \cite{Fischer:2011mz}. 

With or without matter back-coupling to the YM system, we always find
a critical endpoint around $(T^{\text{CEP}},\mu^{\text{CEP}}) \approx
(20\ldots 30, 290)$ MeV. The endpoint is located at low temperatures,
which results from the inclusion of quark and meson fluctuations. In
standard mean-field calculations, where the mesonic fluctuations are
ignored, $T^\text{CEP}$ is typically much higher
\cite{Schaefer:2006ds}.  Already the inclusion of the fermion vacuum
fluctuations lead to much lower $T^{\text{CEP}}$ values, see
e.g.~\cite{Schaefer:2004en,Schaefer:2011ex,Skokov:2010sf}.

Note however, that the present approximation lacks accuracy at large
chemical potential. There we expect baryonic degrees of freedom to be
important. We indeed envisage that it is not so much the baryonic
off-shell fluctuations but rather their importance for the true ground
state that matters. Diquark fluctuations, however, may play a
quantitative r$\hat{\rm o}$le. These considerations are supported by
the findings in two-colour QCD, see e.g.\
\cite{Strodthoff:2011tz,Brauner:2009gu,Ratti:2004ra,Hands:2001ee}.
Hence we conclude that the present approximation lacks predictive
power for $\mu/T\sim 2$. In turn, for smaller ratios it is reliable
and as our computations show, there is no sign of a critical point in this
regime. More details on this matter will
be presented in \cite{HPS}.

%-----------------------------------------------------
%-----------------------------------------------------
\section{Chiral limit}
\label{sec:chirallimit}

\begin{figure}
  \centering
  \includegraphics[width=.49\textwidth]{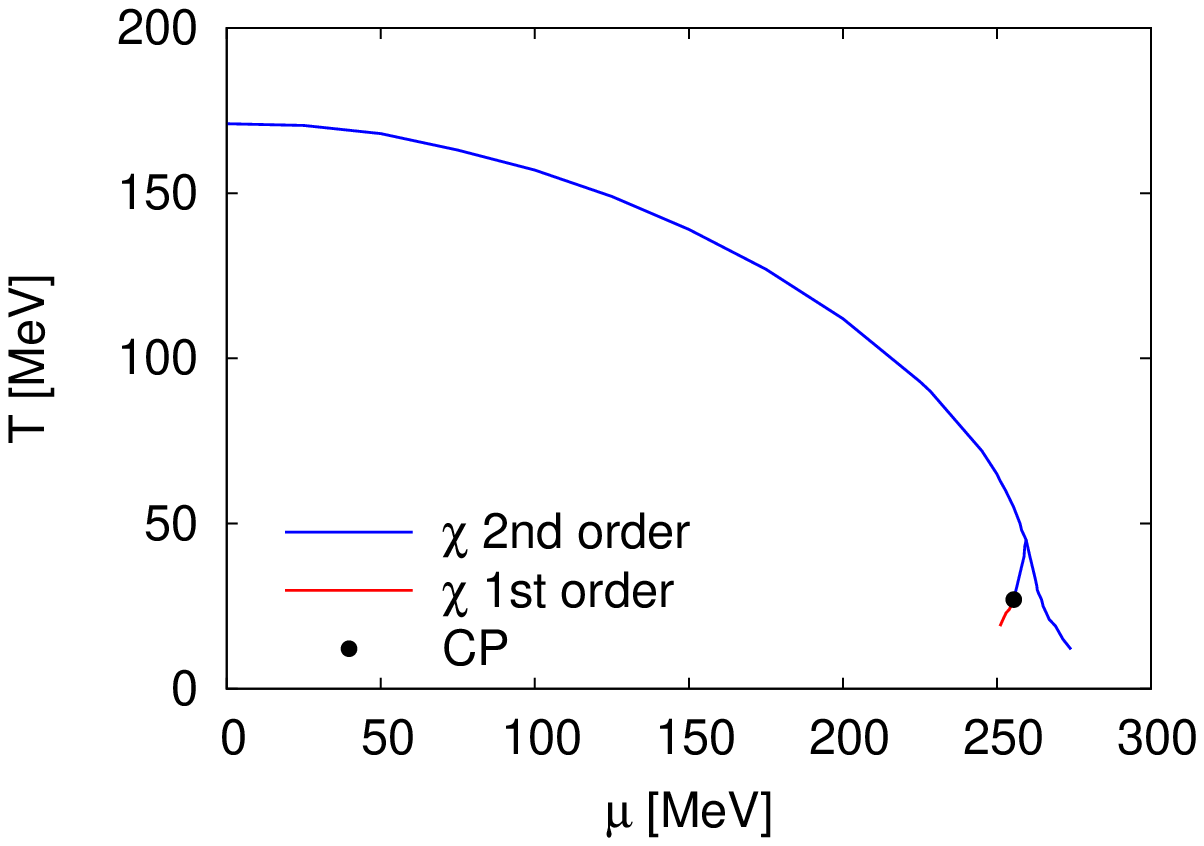}
  \includegraphics[width=.49\textwidth]{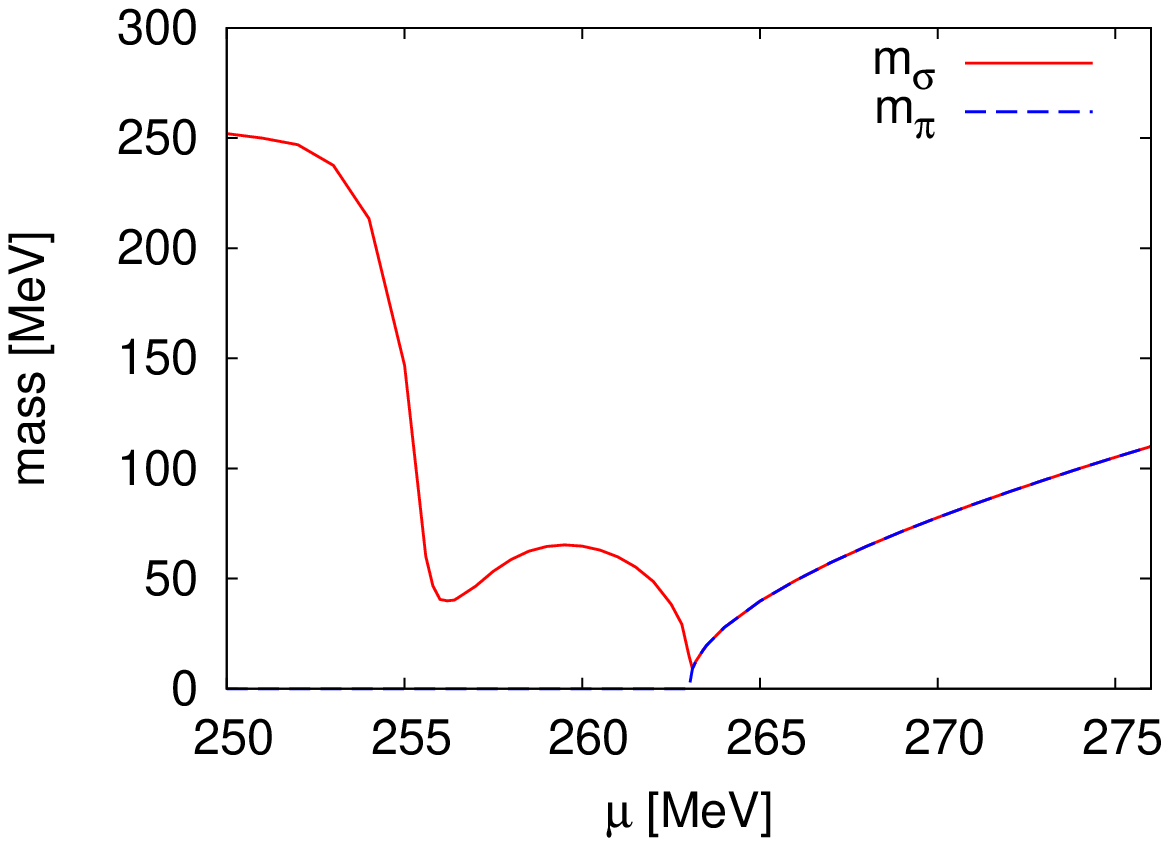}
  \caption{Left: Phase diagram for the chiral transition in the chiral
    limit. Right: Meson masses in the chiral limit at $T=30$ MeV.}
  \label{fig:mpi0}
\end{figure}

In the chiral limit, chiral symmetry is exactly restored at high
temperatures and/or densities and the chiral transition is a phase
transition and not a crossover, see Fig.~\ref{fig:mpi0} (left panel).
At $\mu=0$ we encounter a second-order phase transition, in agreement
with $O(4)$-universality arguments \cite{Pisarski:1983ms}. At large
chemical potential, we find an interesting phase separation: the
second-order chiral transition splits into two branches for decreasing
temperatures. The inner branch (at smaller chemical potential) shows
the inward-bending behaviour characteristic for studies including
thermal and quantum fluctuations. It turns into a first-order
transition for smaller temperatures with a critical endpoint.
The outer branch bends outwards and stays of second-order. This
behaviour is in agreement with previous FRG findings in the pure
quark-meson model in the chiral limit,
cf.~\cite{Schaefer:2004en}. Interestingly, the critical temperature of
the endpoint $T^{\rm CEP}(\mu^{\rm CEP})$ is almost independent of the explicit
symmetry breaking, i.e., independent of the pion mass. This is in
contrast to $T^{\rm CEP}(\mu=0)$ which for decreasing pion mass also
decreases.  Hence, the endpoint lies in the same temperature range as
for physical pion masses. In turn, the critical chemical potential
$\mu^{\rm CEP}$ changes with decreasing pion mass to smaller values. This can
be understood as follows: the chemical potential at which the phase
transition line hits the $\mu$-axis is related to the value of the
quark mass and also to the sigma meson mass. Both masses are smaller
in the chiral limit, cf. also \cite{Schaefer:2008hk}.

The transition splitting is accompanied by two minima in the sigma
meson mass as a function of the chemical potential. This is shown in
Fig.~\ref{fig:mpi0} (right panel) for a temperature above the
endpoint. In contrast to the sigma meson mass, the pions stay massless
until chiral symmetry is completely restored, which happens at the
outer transition branch. For larger chemical potential, both meson
masses degenerate due to the restored chiral symmetry.

\section{Small pion masses}
\label{sec:mpi50}
\begin{figure}
    \centering
    \includegraphics[width=.49\textwidth]{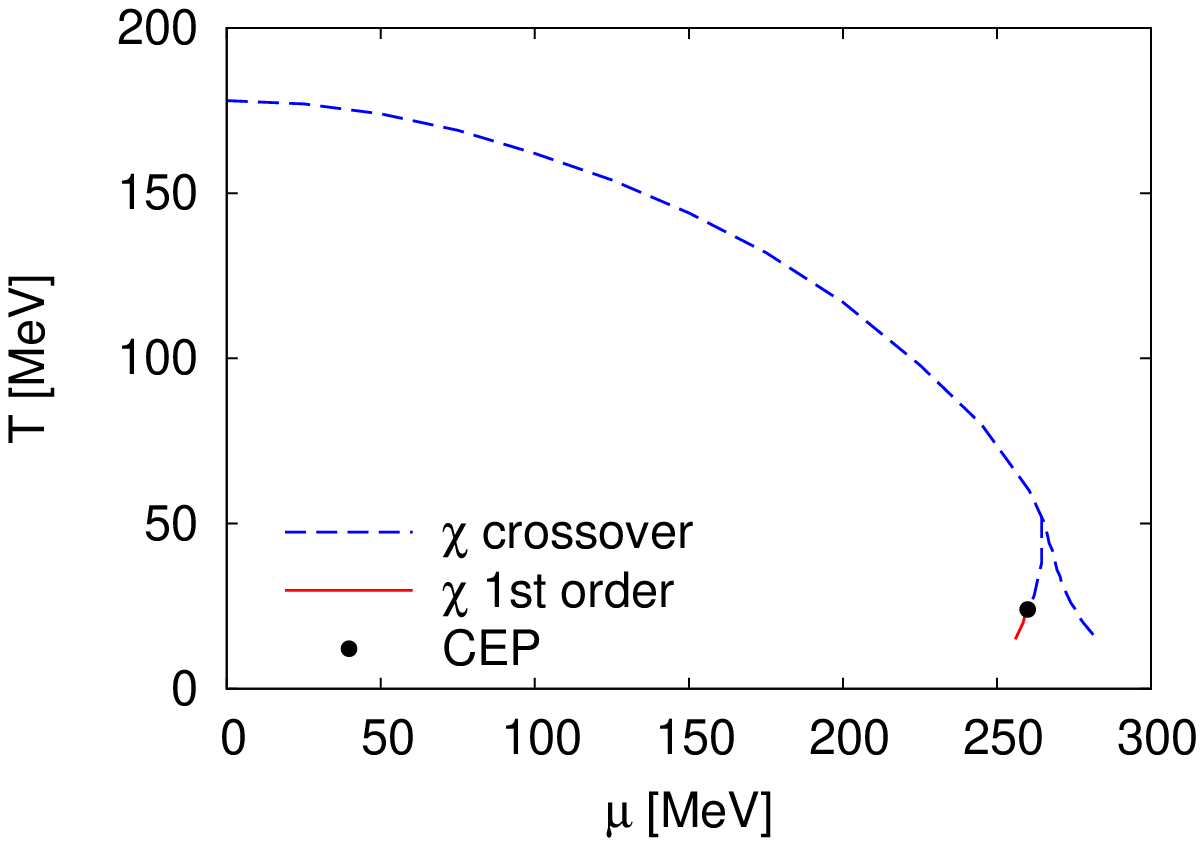}
    \includegraphics[width=.49\textwidth]{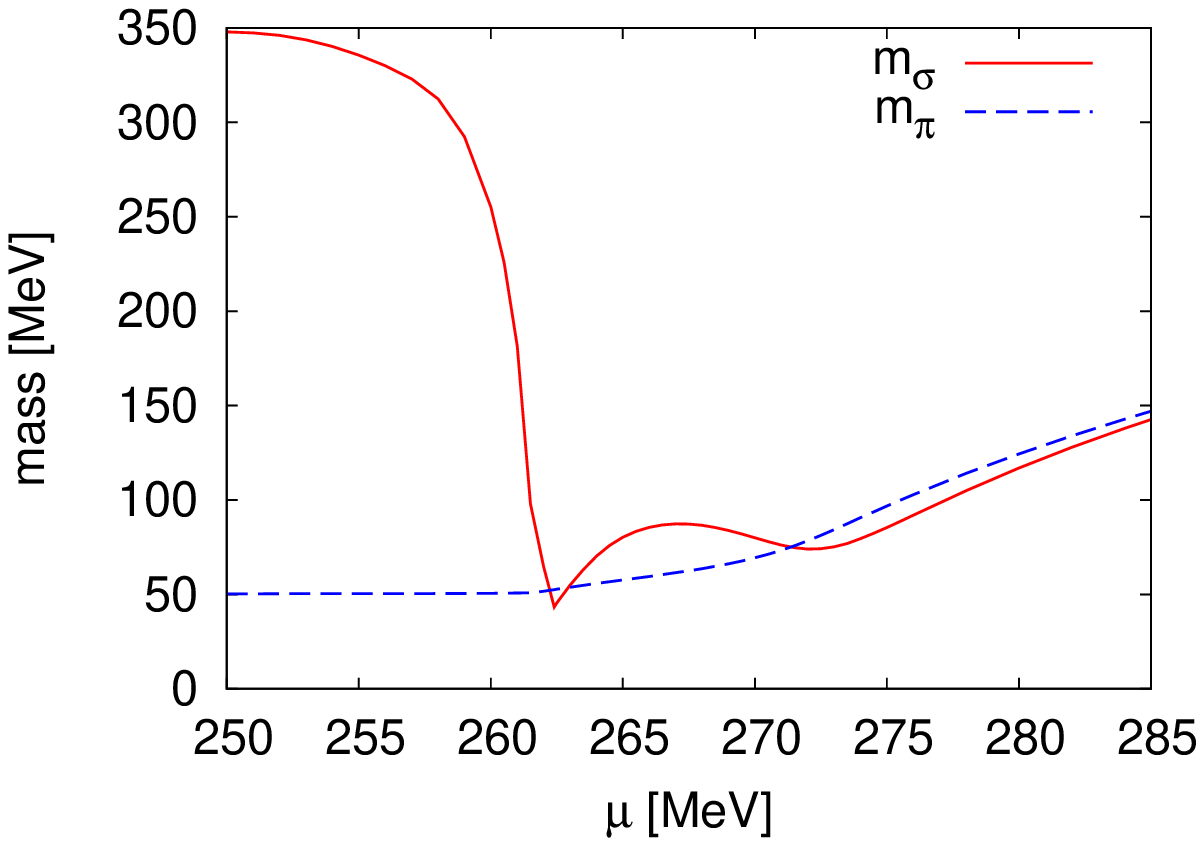}
    \caption{Similar to Fig.~\ref{fig:mpi0}, but for $m_\pi=50$ MeV.}
    \label{fig:mpi50}
\end{figure}
As previously shown, the chiral limit phase structure has peculiar
features at large chemical potential that are not present at
physical pion mass. In the following, we demonstrate how the chiral
limit is connected to the physical mass point and study the influence
of an increasing pion mass on the chiral transition splitting in the
phase diagram.

For a non-vanishing explicit chiral symmetry breaking parameter, the
second-order chiral transition at moderate chemical potential turns
into a crossover.  Apart from this, the phase structure remains
qualitatively the same.  At small pion masses, for example $m_\pi=50$
MeV, the splitting in the chiral transition persists and we find a
critical endpoint on the inner branch, again around $T_c\approx 20-30$
MeV, cf. Fig.~\ref{fig:mpi50} (left). For smaller temperatures the
transition is of first-order. The outer branch remains a crossover for
all temperatures. The sigma meson mass as a function of chemical
potential shows a behaviour similar to the chiral limit. The second
minimum is still visible, but much weaker, see Fig.~\ref{fig:mpi50}
(right). For a larger explicit symmetry breaking this effect is washed
out. Furthermore, the inflection point of the chiral condensate, which
determines the transition line, is also smeared out at larger pion
masses. At physical pion masses no splitting is observed anymore,
cf. Fig.~~\ref{fig:phasediagram_phys}.

\section{Pressure}
\label{sec:TD}

\begin{figure}
    \centering
    \includegraphics[width=.49\textwidth]{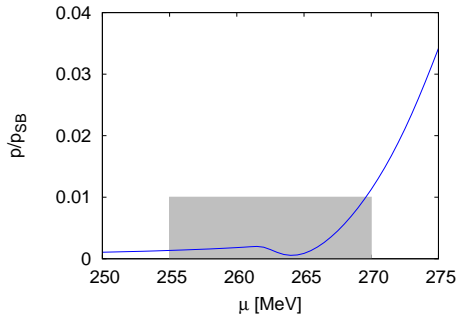}
    \caption{Pressure in the splitting region, $T=30$ MeV,  for $m_\pi= 50$ MeV.}
    \label{fig:T30_TD}
\end{figure}

In order to understand the properties of the chiral splitting region
in the phase diagram in more detail, we also compute the pressure in
this regime. In Fig.~\ref{fig:T30_TD} we show the pressure normalized
by its Stefan-Boltzmann value as a function of $\mu$ for $m_\pi=50$
MeV at the fixed temperature $T=30$ MeV. Clearly, the pressure in the
splitting regime is not a monotonically rising function. A local
minimum in the scaled pressure appears at about the chemical potential
where the sigma meson mass has its local maximum. Only for chemical
potential beyond the second transition branch the pressure increases
significantly. Note however, that the dropping of the pressure for
these densities is unphysical and hints at some shortcoming of the
Polyakov-loop implementation in PQM/PNJL models. As already pointed
out above, the coefficients of the effective Polyakov-loop potential
are fitted to lattice data at vanishing chemical potential.  Some
density aspects of the matter back-coupling are included by our
modifications where $T_0$ has been replaced by $T_0(N_f,\mu)$, but a
fully dynamical treatment within the FRG is still missing.

Indeed, the slope of the used effective Polyakov-loop potential ansatz
is known to be steeper than the corresponding potential in full QCD,
see \cite{Braun:2009gm,Haas:2010bw,Pawlowski:2010ht} which might
result in such an unphysical behaviour. In any case we observe that a
small change in the parameters of the Polyakov loop potential lead to
qualitative changes in the regime under discussion. Hence, in order to
get a more realistic description of the QCD phase diagram, in
particular in this density regime, the full dynamics of the glue
potential $\Omega_{\rm glue}$ has to be taken into account. In our
opinion this, together with the inclusion of baryonic and diquark
degrees of freedom already discussed above, will give us access
to the highly interesting nuclear matter regime at low temperature and
relatively large densities.

%-----------------------------------------------------
%-----------------------------------------------------
\section{Conclusion and Outlook}
\label{sec:conclusion}

In the present contribution we have discussed the effects of quantum
and thermal fluctuations as well as the pion mass sensitivity of the
phase structure of QCD. This is done within the dynamical
Polyakov-extended quark-meson model where fluctuations are included
via the functional renormalization group. We have also argued that the
PQM model can be understood as a well-controlled approximation to
first principle QCD. The precise knowledge of the approximations
involved can be used to systematically improve the PQM model towards
full QCD by successively fixing the model parameters with QCD
input. Specifically, we have considered the back-coupling of quarks to
the glue sector of QCD which results in a $N_f$- and $\mu$ dependent
modification of the $T_0$ parameter in the Polyakov-loop potential.

At physical pion masses, the FRG phase diagram exhibits a critical
endpoint whose location is pushed towards large chemical potential and
low temperature in comparison with a standard mean-field
approximation.  This emphasizes the significant influence of
fluctuations.  We could rule out the emergence of an endpoint for
small chemical potential below $\mu/T\approx 2$.

In the chiral limit, a splitting of the chiral transition line at
large chemical potential and below $T < 50$ MeV could be confirmed
within our truncation. Two transition branches with different
transition orders emerge for small temperatures. On the inner branch,
a second-order critical endpoint as the endpoint of the first-order
transition line could be determined. Within the splitting region
chiral symmetry is still broken.

Interestingly, the area of the splitting region is not very sensitive
to the pion mass. For a small pion mass the splitting phenomenon still
persists and the critical temperature of the endpoint is almost pion
mass independent.  For non-vanishing pion masses the chiral transition
changes from the chiral $O(4)$ universality to a crossover.  Since
fluctuations weaken the chiral phase transition the crossover is in
addition washed out and disappears when we approach physical pion
masses. However, some remnants of the second transition branch are
still present in the chiral condensate at physical pion masses.

The investigation of the thermodynamics in the splitting region might
shed more light on the nature of the different transitions and the
emergence of a critical endpoint. Concerning the pressure in this
region we found that the Polyakov-loop potential has some deficiencies
at low temperature and large chemical potential. Within the splitting
region, the slope of the normalized pressure as a function of 
chemical potential becomes negative, which is an unphysical
behaviour and hampers the computation of other thermodynamic quantities
there. However, in this regime of the phase diagram baryonic degrees
of freedom are of importance but have been neglected so far. 

Furthermore, the inclusion of the matter back-coupling to the glue
sector favours the coincidence of chiral and Polyakov-loop transition
lines also at larger chemical potential. This is in clear
contradistinction to the mean-field results where one observes a
separation of the two transitions. This, together with large-$N_c$
investigations, has been used as an indication for a quarkyonic
phase, where chiral symmetry is restored but matter still
confined. Given the intricacies related with the interpretation of the
Polyakov loop as an order parameter for the confinement/deconfinement
phase transition at finite density, we merely note that our results
with fluctuations cast some doubt on the support provided by the mean-field
 investigations with Polyakov-extended models. Seemingly, it is
the lack of matter back-coupling in these investigations which
triggers the separation of the transition lines. Note also that this
missing back-coupling is built-in in the large-$N_c$ limit. 

In summary, the dynamical Polyakov-loop extended quark-meson model
provides a good approximation to full QCD at small densities. For
densities beyond $\mu/T>2$ the present approximation has to be
extended to include baryonic degrees of freedom, the effects of
in-medium propagation and that of multi-scatterings in a dense medium.

%-----------------------------------------------------
\section*{Acknowledgements}
TKH is recipient of a DOC-fFORTE-fellowship of the Austrian Academy of
Sciences and supported by the FWF doctoral program DK-W1203-N16. JMP
acknowledges support by Helmholtz Alliance HA216/EMMI and BJS by the
Helmholtz Young Investigator Group No. VH-NG- 332.

\bibliographystyle{bibstyle}
\bibliography{bib}

%%%%%%%%%%%%%%%%%%%%%%%%%%%%%%%%%%%%%%%%%%%%%%%%%%%%%%%%%%%%%%%%%%%%%%%%%%%%%%%%%
%%%%%%%%%%%%%%%%%%

\end{document}